# Estado Actual de la Práctica de la Ingeniería de Software en México

Reyes Juárez-Ramírez[1], Karen Cortés Verdín[2], Beatriz Angélica Toscano de la Torre[3], Hanna Oktaba[4], Carlos Alberto Fernández-y-Fernández[5], Brenda Leticia Flores Ríos[6], Fabiola Angulo Molina[7]

[1]*Universidad Autónoma de Baja California, Facultad de Ciencias Químicas e Ingeniería, Tijuana, Baja, California, México;*
[2]*Universidad Veracruzana, Facultad de Estadística e Informática, Xalapa, Veracruz, México;*
[3,7]*Universidad Autónoma de Nayarit, Unidad Académica de Economía, Tepic, Nayarit, México;*
[4]*Universidad Nacional Autónoma de México, Facultad de Ciencias, México D.F., México;*
[5]*Univerversidad Tecnológica de la Mixteca, Instituto de Computación, Huajuapan de León, Oaxaca, México;*
[6]*Universidad Autónoma de Baja California, Instituto de Ingeniería, Mexicali, Baja California, México;*

[1]*reyesjua@uabc.edu.mx*, [2]*kcortes@uv.mx*, [3]*angelica.delatorre@uan.edu.mx*, [4]*hanna.oktaba@ciencias.unam.mx*, [5]*caff@mixteco.utm.mx*, [6]*brenda.flores@uabc.edu.mx*, [7]*fabiola.uan.2010@gmail.com*

**Resumen**

*La ingeniería de software es una disciplina relativamente nueva comparada con otras ciencias, ya que los orígenes del término como tal se remontan a los años 1968 y 1969. En la actualidad, el mercado y la industria del software tienen un auge considerable en varios países del mundo; sin embargo, aunque México se encuentra inmerso en esta carrera, no ha logrado aún el nivel de éxito que se tiene en otros países en este tenor. En este artículo se presenta un panorama sobre la situación que guarda la práctica de la ingeniería de software en México, haciendo énfasis en el ámbito académico. Se muestra un compilado de la actividad de investigación científica realizada en las universidades, así como un análisis breve de los programas educativos de nivel licenciatura que incluyen la disciplina de ingeniería de software. Al final del mismo se plantea el trabajo futuro a realizar para encontrar un punto de convergencia entre la academia y la industria, y poder apoyar así al florecimiento de esta actividad empresarial que tendrá de alguna manera un impacto positivo en la economía de nuestro país.*

## 1. Introducción

El término ingeniería del software (IS) empezó a usarse para expresar el área de conocimiento que se estaba desarrollando en torno a las problemáticas que ofrecía el software en la década de 1960 [4]; este término surge en 1968 durante la primera conferencia [1] realizada por el "NATO Science Committee" y fue reforzado durante la segunda conferencia realizada en 1969 [2], a las que asistieron líderes en el campo de la construcción de sistemas, tanto investigadores como practicantes de la industria.

Sobre la IS varias definiciones han sido propuestas; sin embargo, las más aceptadas son las que a continuación se indican. Boehm adapta la definición de *ingeniería* [3] al contexto del software, diciendo que "*es la aplicación de la ciencia y las matemáticas a través de la cual las propiedades del software son útiles para los clientes*" [4]. Una definición de corte práctico acorde a la disciplina es integrada a partir de tres definiciones previas [5, 6, 7], y dicta que la IS es la aplicación de un enfoque sistemático, disciplinado y cuantificable al diseño, desarrollo, operación y mantenimiento del software, así mismo incluye el estudio de estos enfoques; esto es, la aplicación de la ingeniería a la creación de software.

Aunque la IS es una disciplina relativamente nueva, en 2006 Boehm hace un diagnóstico de cómo se ha desarrollado por décadas [4], cerrando su exposición con las tendencias en tecnologías y la relación que ha de tener la IS con el desarrollo de esa tecnología innovadora. Si bien es cierto que las prácticas de la IS tradicionalmente han sido enfocadas a la generación de sistemas típicos, Boehm señala las tendencias en tecnología que se vislumbran en las décadas venideras [4], hecho que nos debe conducir a reflexionar sobre cómo abrir el panorama de la IS para atender el desarrollo de esas tecnologías, estando involucrada ampliamente la interdisciplinariedad con otras áreas, lo cual significa que la IS debe crear nuevas líneas para cubrir la construcción de este tipo de sistemas.

La IS tiene un sentido muy práctico, incluso los resultados de las investigaciones en el ámbito académico presentan ese perfil de aplicabilidad en la práctica. Debido a estos aspectos, la IS debería tener una amplia aceptación en la industria, aplicando los resultados de las investigaciones que emergen del ámbito académico, así como el resultado de innovación que generan las grandes compañías sobre buenas prácticas en el desarrollo de software. Esto puede dar soporte a las empresas para estar dentro del mercado del software que se expande a pasos considerables. Actualmente, el mercado de software está muy globalizado, resultando más rápido en expansión que otros mercados en desarrollo. Según algunos sitios de Internet, el mercado mundial de software superó los $265 mil millones de dólares en el año 2010, y el crecimiento de este mercado se espera que exceda de 6% anual entre 2010 y 2015, con lo que el mercado llegará casi a $357 mil millones de dólares [10].

Para muchos países el desarrollo de software ha venido a representar un factor determinante en la economía, en la investigación y en la academia; ejemplos de ellos son India, Irlanda, e Israel [8], China, Corea del Sur, y Finlandia [15], así como España, mencionando también a Estados Unidos, Japón y algunos miembros de la Unión Europea [15]. Para el caso de Latinoamérica, países como Brasil, Argentina, Colombia [9], Chile [11], Costa Rica y Uruguay están haciendo sus intentos, siendo Brasil el más avanzado en este tenor [12]. México ha intentado entrar en este contexto, transitando hacia un modelo de desarrollo basado en la innovación. Para lograr este propósito ha apostado por la incursión en el desarrollo de las tecnologías de la información, en especial en la industria del software.

En México la industria de software se consideró un sector estratégico en el sexenio 2000-2006, y varias iniciativas surgieron, siendo la más significativa el programa PROSOFT [13]; sin embargo, el nivel de desarrollo no se ha equiparado con el nivel de competitividad en este rubro que han mostrado otros países. El programa PROSOFT tuvo como metas para el 2013 *"una facturación de 15 mil millones de dólares, aportar el 2.3% del PIB, mantener a la industria como la número uno en América Latina y quinto lugar a nivel mundial, y generar 625,000 empleos"* [16]. En recientes fechas de 2013 se ha afirmado que el mercado de software mexicano expandió su valor a 2 mil 100 millones de dólares respecto al año anterior, lo que significó un crecimiento del 8.5% [17]. Por otro lado, se espera que a finales de 2013 tenga un crecimiento de 8.9%, que equivaldrá a una ganancia de 2 mil 300 millones de dólares; de estos, el 47% corresponderá al mercado de aplicaciones, el 29% al software de implementación y aplicaciones (base de datos, inteligencia de negocios y servidores para entrega), mientras que el 24% corresponderá al software de infraestructura o el relacionado con el hardware (que implica lo relativo a antivirus, sistemas operativos y de virtualización).

Como puede observarse en los apartados anteriores, México produce software y se otorgan servicios con este, sin embargo se desconoce cómo se hace, es decir, no se tiene claro si se desarrolla software apegándose al cuerpo de conocimientos de la disciplina. Debido a su juventud como actividad industrial, el desarrollo de software adopta una forma de organización distinta de la que prevalece en el modelo organizacional tradicional que ha dominado gran parte de la historia industrial [15]; por lo tanto, es conveniente reforzar el desarrollo de software con fundamentos sólidos basados en conocimientos y en buenas prácticas [14]. Dada la desarticulación existente, ya mencionada, entre la industria y la academia, es imperante un trabajo arduo en donde se unifiquen los esfuerzos entre estos dos sectores para lograr una mejor práctica de esta disciplina.

El software es una industria muy atractiva, ya que es intensiva en conocimiento y en mano de obra calificada; sin embargo, requiere una fuerte disciplina de los países para contar con la fuerza laboral que exige la industria, por lo que para las economías emergentes como México el reto es ajustar la capacitación del recurso humano al desarrollo de la industria [15]. Para lograr un surgimiento exitoso de esta industria, es conveniente realizar un análisis de las capacidades y las vocaciones de los actores involucrados: la industria, la academia y el gobierno.

Para dar inicio a este estudio, en este artículo se presenta una visión de la productividad en el ámbito académico, en cuanto a las actividades de investigación y prácticas de la IS. En futuras publicaciones se presentará un panorama de la industria, así como la convergencia que hay y que debe haber entre la academia y la industria.

Este artículo está organizado como a continuación se indica. La sección 2 contiene un compilado de la actividad de investigación realizada en el ámbito académico, organizada con base en las áreas y sub-áreas del conocimiento de la IS contenidas en el SWEBOK. En esta sección se detecta a cuáles áreas está enfocada la actividad de investigación realizada en varias de las universidades públicas del país, ya que se ha hecho un compendio de las publicaciones realizadas principalmente en los últimos tres años. La sección 3 presenta un breve compilado de los programas educativos relacionados con la IS, enfatizando en el planteamiento de un análisis a realizar en el futuro sobre la cobertura de las áreas del SWEBOK para preparar dentro de las universidades al recurso humano calificado en la práctica de la IS. La sección 4 contiene las conclusiones y el trabajo futuro a realizar en el contexto mexicano dentro de la disciplina de la IS.

## 2. Investigación en Ingeniería de Software dentro del Ámbito Académico

En esta sección se describe la actividad de investigación realizada en el ámbito académico en México, tomando una muestra de nueve grupos de investigación a lo largo del país. Se toma como base el cuerpo de conocimientos en ingeniería de software SWEBOK v3.0 [18], que es un incremento de la versión 2004 contenida en el "Technical Report ISO/IEC TR 19759" [19], y que tiene como propósitos: 1) caracterizar los contenidos de la disciplina de ingeniería de software; 2) promover una vista consistente de la ingeniería de software a nivel mundial; 3) clarificar y establecer los límites de la ingeniería de software respecto a otras disciplinas; 4) proporcionar los fundamentos para materiales de entrenamiento y desarrollo de la curricula y 5) proporcionar las bases para la certificación y el licenciamiento de los ingenieros de software. El SWEBOK contiene 15 áreas de conocimiento (véase Tabla 1), las cuales serán tomadas como base para este compilado.

Se tiene conocimiento de la existencia de varios grupos de investigación en instituciones públicas y privadas en México, tanto universidades como centros de investigación; sin embargo, con motivo de visualizar las actividades de investigación que se realizan, se tomaron como muestra solamente nueve grupos de universidades públicas.

Los nueve grupos de investigación considerados son: Universidad Autónoma de Aguascalientes, Universidad Veracruzana, Universidad Autónoma de Nayarit, Universidad Autónoma de Baja California (UABC) Campus Tijuana y Mexicali, Universidad Nacional Autónoma de México (UNAM), Universidad Autónoma de Querétaro y Universidad Autónoma de Hidalgo.

**Tabla 1**: Áreas consideradas en el SWEBOK

| Área clave |
|---|
| **AC_01.** Economía de la ingeniería de software |
| **AC_02.** Requerimientos de software |
| **AC_03.** Pruebas de software |
| **AC_04.** Construcción de software |
| **AC_05.** Administración de la configuración de software |
| **AC_06.** Fundamentos de computación |
| **AC_07.** Modelos y métodos de la ingeniería de software |
| **AC_08.** Mantenimiento de software |
| **AC_09.** Fundamentos matemáticos |
| **AC_10.** Diseño de software |
| **AC_11.** Administración de la ingeniería de software |
| **AC_12.** Práctica profesional de la ingeniería de software |
| **AC_13.** Fundamentos de ingeniería |
| **AC_14.** El proceso de la ingeniería de software |
| **AC_15.** Calidad de software |

Para una mejor ubicación de las publicaciones, las diferentes áreas del SWEBOK fueron desglosadas en temas y sub-temas; sin embargo, por cuestiones de espacio en las siguientes tablas solo se indica el tema (véanse Tablas 2-11). Para hacer este compilado se consideró la producción principalmente de los últimos tres años, y tomando solamente a lo más 20 publicaciones por grupo de investigación; lo cual significa que hay más investigación producida. Las publicaciones fueron ubicadas en los temas de cada área con base en cómo cubren los subtemas (para más detalles los subtemas consultar [18]).

Algunas áreas que no tienen mucha cobertura de investigación no son mostradas en tablas, pero se describen a continuación. Del área "Mantenimiento de software" se investiga en los temas "Fundamentos del mantenimiento de software" [54, 62] y "Herramientas de Software para el mantenimiento" [54, 62]. Del área "Fundamentos matemáticos" se investiga en los temas "Conjuntos, relaciones, funciones" [20], "Técnicas de prueba" [20], "Fundamentos de conteo" [20, 37], "Grafos y árboles" [20], "Gramáticas" [22, 29, 34]. Del área "Economía de la ingeniería de software" se investiga en el tema "Fundamentos de la economía de la ingeniería de software" [64, 78, 82, 96, 97, 129]. Del área "Administración de la configuración de software" se investiga en el tema "Herramientas de administración de la configuración del software" [54, 62]. Sobre el área "Pruebas de Software" se investiga en todos sus temas [76].

**Tabla 2**: Requerimientos de software

| Tema | Publicaciones |
|---|---|
| **AC_02_01.** Fundamentos de requerimientos de software. | 20, 22, 34, 46, 106, 114, 124, 129 |
| **AC_02_02.** Proceso de requerimientos. | 64, 72, 74, 106, 114, 124 |
| **AC_02_03.** Elicitación de requerimientos. | 106, 114, 124 |
| **AC_02_04.** Análisis de requerimientos. | 22, 30, 87, 106, 114, 124 |
| **AC_02_05.** Especificación de requerimientos. | 20, 22, 30, 85, 87, 90, 91, 92, 94, 106, 114, 124 |
| **AC_02_06.** Validación de | 22, 30, 85, 87, 90, 91, |
| requerimientos. | 92, 94, 106, 114 |
| **AC_02_07.** Consideraciones prácticas. | |
| **AC_02_08.** Herramientas para requerimientos de software. | 22, 29, 30, 34, 86, 91, 94 |

**Tabla 3**: Construcción de software

| Tema | Subtema |
|---|---|
| **AC_04_01.** Fundamentos de la creación de software | |
| **AC_04_02.** Administración de la construcción | 25, 26, 75, 81 |
| **AC_04_03.** Consideraciones prácticas | |
| **AC_04_04.** Tecnologías de construcción | 51, 78 |
| **AC_04_05.** Herramientas de construcción de software | 55 |

**Tabla 4**: Fundamentos de computación

| Tema | Subtema |
|---|---|
| **AC_06_01.** Técnicas de solución de problemas | 55, 108 |
| **AC_06_02.** Abstracción | 89, 93, 108 |
| **AC_06_03.** Fundamentos de programación | 89, 93, 108 |
| **AC_06_04.** Bases de lenguajes de programación | 55, 73 |
| **AC_06_05.** Técnicas y herramientas de "depuración" | |
| **AC_06_06.** Estructuras de datos y representaciones | 108 |
| **AC_06_07.** Complejidad y algoritmos | |
| **AC_06_08.** Conceptos básicos de un sistema | 51 |
| **AC_06_09.** Factores básicos de usuarios humanos. | 24, 23, 39, 40, 48 |
| **AC_06_10.** Factores básicos del desarrollador humano. | |

**Tabla 5**: Modelos y métodos de la ingeniería de software

| Tema | Subtema |
|---|---|
| **AC_07_01.** Modelado | 52, 85 |
| **AC_07_02.** Tipos de modelos | 49 |
| **AC_07_03.** Análisis de modelos | 52, 60, 67 |
| **AC_07_04.** Métodos de ingeniería de software | 20, 21, 24, 27, 35, 67, 69, 71, 85, 90, 92, 94 |

**Tabla 6**: Diseño de software

| Tema | Subtema |
|---|---|
| **AC_10_1.** Fundamentos del Diseño de Software | 101, 102 |
| **AC_10_2.** Principales problemáticas del diseño de Software | 128 |
| **AC_10_3.** Estructura y arquitectura del Software | 23, 33, 36, 75, 101, 102, 103, 104, 111, 128 |
| **AC_10_4.** Diseño de interfaz de usuario | 23, 24, 31, 32, 33, 36, 38, 39, 40, 48, 55, 115, 116, 118, 121, 122, 123, 127 |
| **AC_10_5.** Análisis y evaluación de la calidad del diseño de software | 23, 101, 102, 103, 104, 111, 115, 116, 118, 121 |
| **AC_10_6.** Notación en el diseño de software | 85, 86, 91, 94, 121, 128 |
| **AC_10_7.** Estrategias y métodos en el diseño de software | 31, 32, 84, 101, 102, 103, 104, 111, 118 |
| **AC_10_8.** Herramientas para el diseño de software | 86, 91, 94 |

**Tabla 7**: Administración de la ingeniería de software

| Tema | Subtema |
|---|---|
| **AC_11_01.** Iniciación y definición de alcance | |
| **AC_11_02.** Planeación del proyecto de software | 57, 59, 96, 97, 119 |
| **AC_11_03.** Promulgación del proyecto de software | 96, 97, 119 |
| **AC_11_04.** Revisión y evaluación | 111, 119 |
| **AC_11_05.** Cierre | |
| **AC_11_06.** Medición de la ingeniería de software | 111 |
| **AC_11_07.** Herramientas de medición de software | 20, 25, 26, 71 |

**Tabla 8**: Práctica profesional de la ingeniería de software

| Tema | Subtema |
|---|---|
| **AC_12_01.** Profesionalismo | 38, 47, 50, 56, 61, 54, 63, 100, 105, 107, 125 |
| **AC_12_02.** Dinámica y psicología de grupos | 46, 100 |
| **AC_12_03.** Habilidades de comunicación | 26, 25, 38, 46, 107 |

**Tabla 9**: Fundamentos de ingeniería

| Tema | Subtema |
|---|---|
| **AC_13_01.** Técnicas experimentales y métodos empíricos | 20, 21, 22, 58, 62, 73 |
| **AC_13_02.** Análisis estadístico | 20, 21, 22, 62 |
| **AC_13_03.** Mediciones | 21, 62, 74 |
| **AC_13_04.** Diseño en ingeniería | 51 |
| **AC_13_05.** Modelado, prototipos, simulaciones | 21, 63 |
| **AC_13_06.** Estándares | 24, 47, 48, 50, 56, 59, 61, 63, 72, 76, 83 |
| **AC_13_07.** Análisis de causas raíz. | 21, 89 |

**Tabla 10**: El proceso de la ingeniería de software

| Tema | Subtema |
|---|---|
| **AC_14_01.** Definición del proceso de software | 26, 31, 32, 53, 73, 95, 98, 105, 109, 122, 125 |
| **AC_14_02.** Ciclos de | 28, 60, 72, 73, 80, 82, 83, 84, 95, |
| vida del software | 98, 109, 110, 113, 117, 120, 122 |
| **AC_14_03.** Evaluación y mejora del proceso de software | 64, 65, 66, 70, 74, 75, 76, 14, 78, 79, 80, 110, 113, 117, 120, 125, |
| **AC_14_04.** Medición del software | 21, 64, 65, 66, 70, 74, 75, 76, 77, 78, 79, 80, 110, 113, 117, 120, 129 |
| **AC_14_04.** Herramientas para el proceso de ingeniería de software | 26, 28, 47, 61 |

**Tabla 11**: Calidad de software

| Tema | Subtema |
|---|---|
| **AC_15_01.** Fundamentos de calidad de software | 21, 27, 29, 35, 58, 52, 64, 65, 66, 70, 75, 103, 105, 110, 112, 113, 115, 116, 117, 122, 123, 125, 126, 127, 128 |
| **AC_15_02.** Proceso de administración de la calidad de software | 21, 99, 103, 110, 125, 126, 127, 128 |
| **AC_15_03.** Consideraciones prácticas | 21, 99, 101, 102, 103, 111, 115, 116, 127 |
| **AC_15_04.** Herramientas para la calidad de software | 29, 30, 77, 78, 79, 80, 99, 103 |

Como puede verse en las tablas 10, 6, 11, 4, y 2, las áreas más atendidas en la investigación que realizan las universidades son: El proceso de la ingeniería de software, Diseño de software, Calidad de software, Fundamentos de ingeniería, Requerimientos de software. En siguiente medida son atendidas estas cuatro áreas: Modelos y métodos de la ingeniería de software, Fundamentos de la computación, Práctica profesional de la ingeniería de software y Administración de la ingeniería de software.

## 3. Programas Educativos Relacionados con Ingeniería de Software

En México se tiene oferta de programas a nivel Licenciatura que cubren en cierta medida las áreas de la IS. Una mejor referencia de dichos programas puede encontrarse en los padrones de la Asociación Nacional de Instituciones de Educación en Tecnologías de la Información A.C. (ANIEI). La ANIEI agrupa las áreas de las ciencias de la computación de la siguiente manera [44]: Matemáticas, Arquitectura de computadoras, Redes, Programación e ingeniería de software, Tratamiento de información, Interacción hombre-máquina, Software de base y entorno social. De acuerdo al peso de estas áreas define cuatro dominios de desarrollo profesional en torno a la informática y computación. Los perfiles asociados se resumen en [43]: Licenciado en Informática, Licenciado en Ciencias Computacionales, Ingeniero en Computación e Ingeniero en Software.

A nivel nacional se ofertan programas de nivel licenciatura en el área de las TIC a lo largo del territorio nacional (en todos los Estados), siendo un considerable número de ellos acreditados por el Consejo Nacional de Acreditación de Informática y Computación (CONAIC). De estos programas, 33 son de tipo ingeniería y 55 son de tipo licenciatura [42]. Todos ellos alguna vez acreditados (algunos de los cuales aún tienen vigencia de acreditación), lo que significa que al menos el 17.5 por ciento de su plan de estudios son créditos del área de Programación e Ingeniería de Software, considerando la tabla de áreas que maneja la ANIEI [43]. Es importante señalar que existen otros programas de tipo ingeniería que son acreditados por otros organismos acreditadores como el Consejo de Acreditación de la Enseñanza en Ingeniería (CACEI) [45], y que no están acreditados a la vez por el CONAIC, pero que reúnen también bloques curriculares sobre la IS.

De este conjunto de programas acreditados por ambos organismos, los nombres más comunes son Ingeniería en Sistemas Computacionales, Ingeniería en Computación, Ingeniería en Sistemas de Información, Ingeniería en Tecnologías de la Información, Ingeniería en Computación y Sistemas e Ingeniería en Ciencias de la Computación. Algunos programas tienen nombres más enfocados tales como Ingeniería de Desarrollo de Software e Ingeniería en Informática; todos ellos con un énfasis muy importante en el desarrollo de software. De la misma forma, los programas del tipo licenciatura acreditados por el CONAIC versan con nombres tales como Licenciatura en Informática, Licenciatura en Ciencias Computacionales, Licenciatura en Tecnologías de Información, Licenciatura en Sistemas Computacionales, Licenciatura en Administración de Tecnologías de Información y Licenciado en Sistemas de Información Administrativa.

Existen otros programas educativos con énfasis en ingeniería de software, tal es el caso de Ingeniería en Tecnologías de la Computación y Telecomunicaciones (Universidad Iberoamericana) e Ingeniero en Tecnologías de la Información y Comunicaciones (ITESM), que aunque tienen nombres y orientación genéricos si tienen un área de especialización en la disciplina [41].

Los programas educativos más antiguos en ingeniería de software tienen poco más de 10 años que fueron constituidos [41]; sin embargo, se ha visto un crecimiento moderado en la creación de programas en ingeniería de software no solo a nivel nacional sino mundial en el nivel licenciatura. En México los programas más acordes a la IS son los siguientes:

Licenciado en Ingeniería de Software (Facultad de Matemáticas, Universidad Autónoma de Yucatán), Ingeniería de Software (Universidad Madero, Puebla; CETYS Universidad, Ensenada; Facultad de Informática, Universidad Autónoma de Querétaro) e Ingeniería en Desarrollo de Software (CESUN Universidad, Tijuana).

De manera general, a nivel nacional y mundial, el término "Ingeniería de Software" todavía no es de amplio uso a nivel licenciatura como nombre de un programa educativo, y suele sustituirse por términos más comunes como Ingeniería en Desarrollo de Software, Ingeniería en Informática, Ingeniería en Sistemas Computacionales e incluso se usa el término de Ingeniería en Computación con un énfasis muy importante en el desarrollo de software. Por otro lado, parece haber un mayor interés en programas de nivel posgrado en esta área, como especializaciones, maestrías y doctorados, evidencia de ello es que la investigación realizada en el ámbito académico se realiza dentro del contexto de posgrado; sin embargo, tampoco se usan nombres precisos en Ingeniería de Software para referir a tales programas.

Particularmente en México, para cumplir con las metas del programa PROSOFT en cuanto a la formación de recurso humano calificado de base, es conveniente contar con un profesional en desarrollo de software integral en la disciplina, formación que pueda ofertarse en diferentes universidades, no solo en aquellas universidades más flexibles y sensibles a los requerimientos del mercado laboral y a la oferta de nuevas opciones educativas. Con base en lo anterior, es conveniente realizar un análisis más detallado sobre la cobertura del cuerpo de conocimientos de la IS que los programas educativos cubren.

## 4. Conclusiones y Trabajo Futuro

En este artículo se exploran brevemente los orígenes y definiciones de la IS. De la misma forma, se contextualiza la importancia del mercado de software en la actualidad y por consecuencia la importancia de la práctica de la IS. Se plantea que en México hace falta conocer las capacidades y prácticas tanto de la academia como de la industria, esto con el fin de poder definir una vocación para el país, misma que apoye el surgimiento de una prominente industria del software. Se presenta un compilado de la actividad de investigación en el ámbito académico, en términos de producción científica sobre la disciplina de ingeniería de software. Este compilado permite ver las áreas del cuerpo de conocimientos que son más atendidas en esta actividad de investigación. De la misma manera, se presenta un breve compilado de los programas educativos a nivel licenciatura que incluyen áreas de especialización en ingeniería de software.

El compilado realizado sobre la investigación que se hace en las universidades permite detectar las vocaciones de cada grupo. Por ejemplo, el grupo UABC Tijuana presenta una vocación en "Requerimientos de software", "Diseño de software" y "Calidad de software", con matices de "Fundamentos matemáticos" y "Fundamentos de ingeniería". El grupo UABC Mexicali presenta una vocación en las áreas de "Fundamentos de ingeniería", "Práctica profesional de la ingeniería de software" y "Modelos y métodos de la ingeniería de software". El grupo de la Universidad Autónoma de Aguascalientes presenta una vocación en "Requerimientos de software", "Diseño de software", "El proceso de ingeniería de software" y "Calidad de software". El grupo de la UNAM presenta una fuerte vocación en las áreas del "Proceso de ingeniería de software" y "Calidad de software". El grupo de la Mixteca presenta una vocación en "Requerimientos de software", "Diseño de software" y "Modelos y métodos de la ingeniería de software". El grupo de la Universidad Veracruzana presenta una vocación en "Diseño de software" y "Calidad de software". El grupo de la Universidad Autónoma de Tepic presenta inicios en el área de "Práctica profesional de la ingeniería de software". De igual manera, el grupo de la Universidad Autónoma de Querétaro muestra inicios en las áreas de "El proceso de la ingeniería de software", "Administración de la ingeniería de software" y "Calidad de software". Finalmente, el grupo de la Universidad Autónoma de Hidalgo presenta inicios en el área de "El proceso de la ingeniería de software".

El material mostrado en este artículo representa apenas el inicio de una investigación que dará como resultado tener un mapa de las actividades en ingeniería de software realizadas tanto en la academia como en la industria. El trabajo futuro se plantea de la siguiente manera:

   a) Realizar un estudio de las prácticas en ingeniería de software que realiza la industria de desarrollo de software en México. Este estudio permitirá ver los diferentes niveles de empresas que hay (grandes, medianas, pequeñas) y las prácticas en cada nivel de empresas. Esto permitirá conocer sus vocaciones actuales.
   b) Realizar un cruce entre las actividades de investigación en la academia y las prácticas en la industria. Esto permitirá encontrar convergencias y/o proponer ajustes a realizar para una mejor articulación.
   c) Hacer propuestas de ajustes a realizar en los perfiles que deben cubrir los programas

educativos, con el fin de cubrir de mejor manera las necesidades de las prácticas en la industria, pero sin dejar de lado el cuerpo de conocimientos de la ingeniería de software.

Finalmente, es importante realizar otro análisis sobre los programas de posgrado orientados a ingeniería de software, para detectar las áreas del cuerpo de conocimientos que están descubiertas y proponer acciones para cubrirlas, ya que es conveniente generar conocimiento en todas las áreas y sobre todo, visualizar la forma de poner en práctica las investigaciones dentro del contexto de la industria.

## 5. Agradecimientos

En esta sección se expresan los agradecimientos a los colaboradores para la realización de este artículo.

Este artículo no hubiera podido ser realizado sin la colaboración de los coautores, quienes proporcionaron información de sus respectivos grupos de investigación. Su participación es un ejemplo de que unidos podemos sacar adelante iniciativas que beneficien a la comunidad de ingeniería de software en México y al país en general. Por otro lado, un agradecimiento especial a los estudiantes del programa Maestría y Doctorado en Ciencias e Ingeniería (MyDCI) en UABC Tijuana (área: ciencia de la computación), por su valiosa colaboración para dar formato a las referencias; estoy seguro que esta actividad dejó una reflexión en ellos. Finalmente, un agradecimiento especial a los estudiantes de los cursos Ingeniería de Software y Aseguramiento de Calidad de Software del ciclo 2013-2 en UABC Tijuana, por inspirarme a presentarles un panorama de la situación actual de México en esta disciplina, en la que varios de ellos realizarán su carrera profesional.

## 6. Referencias